\documentclass[final,preprint,draft,showpacs,tightenlines,aps,fleqn]{revtex4}% 
\usepackage{graphicx}
\usepackage{bm}
\usepackage{dcolumn}
\usepackage{amsmath}%
\usepackage{amsfonts}%
\usepackage{amssymb}
\begin{document}

\begin{flushright}
PNU-NTG-07/2003\\
TPJU-10/2003 \\
BNL-NT-03/37
\end{flushright} 
\vspace*{1cm}  
\title{Magnetic moments of exotic pentaquarks
in the chiral quark-soliton model}
\author{Hyun-Chul Kim$^{(1)}$\footnote{E-mail
    address:~hchkim@pusan.ac.kr} and Micha{\l} 
Prasza{\l}owicz$^{(2)}$\footnote{E-mail address:~michal@quark.phy.bnl.gov,
Fulbright Fellow on leave from M. Smoluchowski
Institute of Physics, Jagellonian University, Krak{\'o}w, Poland}}
\affiliation{$^{(1)}$~Department of Physics, and \\
Nuclear Physics and Radiation Technology Institute (NuRI),
Pusan National University,
609-735 Busan, Republic of Korea,\\
$^{(2)}$~Nuclear Theory Group,
Brookhaven National Laboratory,
Upton, NY 19973-5000 \\~~}
%\date{November 2003}

\begin{abstract}
~~~\\
We investigate the magnetic moments of the baryon antidecuplet within the
framework of the chiral quark-soliton model in the chiral limit in a
``\emph{model-independent}'' approach. Sum rules for the magnetic moments are
derived. The magnetic moment of $\Theta^{+}$ is found to be about
$0.2\sim 0.3\,\mu_{N}$.

\end{abstract}
\maketitle

%%%%%%%%%%%%%%%%%%%%%%%%%%%%%%%%%%%%%%%%%%%%%%%%%%%%%%%%%%%%%%%%%%%%%%%%%%%

%\tableofcontents{}

%\section{Introduction}

\textbf{1.} Recently, the pentaquark ($\mathrm{uudd}\bar{\mathrm{s}}$) baryon
$\Theta^{+}$ was found by LEPS collaboration~\cite{Nakano:bh}, motivated by a
theoretical predicton of the chiral quark-soliton
model~\cite{Diakonov:1997mm}. In fact, exotic SU(3) representations
containing exotic baryonic states are naturally accomodated in the
chiral models~\cite{Chemtob,manohar}, where the quantization condition
emerging from the Wess-Zumino-Witten term selects SU(3)
representations of triality zero~\cite{su3quant}. The early estimate
of the $\Theta^{+}$ mass in the Skyrme
model~\cite{Praszalowicz:2003ik} is in very good agreement with the
present experimental results. 

A series of experiments confirmed the existence of $\Theta^{+}$
~\cite{Barmin:2003vv,Stepanyan:2003qr,Barth:2003es,neutrino}. The
measured mass of $\Theta^{+}$ is about $1.54\,\mathrm{GeV}$ with a very narrow
width below $25\,\mathrm{MeV}$ which is consistent with the prediction of
Ref.~\cite{Diakonov:1997mm}.  In addition to $\Theta^+$, the NA49
collaboration at CERN has recently announced the finding of the new
exotic baryons $\Xi_{3/2}$~\cite{NA49} with strangeness $S=-2$ and
isosipn $T=3/2$, which are also the members of the antidecuplet.     

The discovery of $\Theta^{+}$ has motivated theorists to look more
closely into the possible exotic states in quark models \cite{quark},
chiral models \cite{chiral,chiral2}, lattice QCD \cite{lattice},
nucleon-meson bound states models \cite{bound}, constituent quark
models \cite{Stancu:2003if}, QCD sum rules \cite{Zhu:2003ba,summag}
and in group theoretical approach \cite{groupth}. Some authors
questioned the applicability of the rigid rotator approach to the
exotic states within the chiral models \cite{chiral2,negative}.  For a
comprehensive review of different models, see Ref.\cite{review}. 

The cross sections for the $\Theta^{+}$ production from nucleons induced by
hadrons~\cite{Liu:2003rh,Hyodo:2003th} and
photons~\cite{Liu:2003rh,Nam:2003uf} have been already described
theoretically. However, while it is crucial to know the magnetic moment of
$\Theta^{+}$ in order to study its production via photo-reactions, it is yet
to be determined. The magnetic moments of the baryon octet and decuplet have
been successfully described within the chiral quark-soliton model ($\chi
$QSM)~\cite{Kim:1997ip,Kim:1998gt} in a ``model-independent'' approach. The
``model-independent'' analysis has an advantage over dynamical model
calculations, since it only makes use of the underlying symmetries,
with the experimental data of the octet magnetic moments used as an
input. In Refs.~\cite{Kim:1997ip,Kim:1998gt} we calculated the magnetic
moments of $\Delta^{++}$ and $\Omega^{-}$, finding satisfactory
agreement with the existing data. We have also predicted the magnetic
moment of $\Delta^{+}$ which was measured quite
recently~\cite{Kotulla:2002cg}. 

In this letter, we extend the former investigation~\cite{Kim:1997ip} to the
magnetic moments of the baryon antidecuplet in the chiral limit. In
particular, we find that the magnetic moments of the baryon antidecuplet are
unexpectedly small, similarly to the recent finding of Ref.\cite{summag}.
In addition, we obtain the sum rules for the antidecuplet magnetic moments
similar to the Coleman-Glashow sum rules~\cite{Coleman:1961jn}.

\textbf{2}. In the chiral limit the collective magnetic moment operator can be
parametrized within the framework of the $\chi$QSM as follows:
\begin{equation}
\hat{\mu}=v_{1}D_{Q3}^{(8)}\;+\;v_{2}d_{pq3}D_{Qp}^{(8)}\cdot\hat{J}_{q}%
\;+\;\frac{v_{3}}{\sqrt{3}}D_{Q8}^{(8)}\hat{J}_{3}, \label{Eq:mu}%
\end{equation}
where the dynamical variables $v_{i}$ encode dynamics of the chiral soliton
and are independent of the baryon considered. They are generically expressed
in terms of the inertia parameters of the soliton in the $\chi$QSM:
\begin{equation}
\sum_{m,n}\langle n|\Gamma_{1}|m\rangle\langle m|\Gamma_{2}|n\rangle
\mathcal{R}(E_{n},E_{m},\Lambda), \label{spec}%
\end{equation}
where $\Gamma_{i}$ are spin-isospin operators acting on the quark eigenstates
$|n\rangle$ of the one-body Dirac Hamiltonian in the soliton-background field.
The double sum over all the eigenstates can be evaluated
numerically~\cite{Goeke:fk,Wakamatsu:1990ud,Blotz:1992pw}. Since its sea part
diverges, one requires the regularization expressed by $\mathcal{R}$ with the
cut-off parameter $\Lambda$. However, instead of calculating the
dynamical variables $v_{i}$ numerically, two linear combinations of
them can be fitted to the experimental data of the octet magnetic 
moments~\cite{Kim:1997ip,Kim:1998gt}. $D_{ab}^{(\mathcal{R})}(R)$ denotes the
SU(3) Wigner function, $R(t)$ is the time-dependent SU(3) matrix responsible
for the rotation of the soliton in the collective coordinate space
\cite{su3quant,Blotz:1992pw,Christov:1995vm}. $Q$ is the quark electric
charge operator and $\hat{J}_{a}$ stands for an operator of the generalized
spin acting on the baryonic wave functions $\psi_{B_{\mathcal{R}}}(R)$.

In order to evaluate the magnetic moments of the baryon antidecuplet, we need
to calculate the following matrix elements:
\begin{equation}
\mu_{B_{\overline{10}}}=\int dR\psi_{B_{\overline{10}}}^{\ast}(R)\hat{\mu
}(R)\psi_{B_{\overline{10}}}(R),\label{Eq:matrix_e}%
\end{equation}
where the wave functions $\psi_{B_{\mathcal{R}}}(R)$ take the following form:
\begin{equation}
\psi_{B_{\mathcal{R}}}(R)=\psi_{(\mathcal{R};Y,T,T_{3})(\mathcal{R}^{\ast
};-Y^{\prime},J,J_{3})}=\sqrt{\mathrm{dim}(\mathcal{R})}(-1)^{J_{3}-Y^{\prime
}/2}D_{Y,T,T_{3};Y^{\prime},J,-J_{3}}^{(\emph{R})\ast}(R).\label{Eq:wave_f}%
\end{equation}
Here $\mathcal{R}$ stands for the allowed irreducible representations of the
SU(3) flavor group, \emph{i.e.} $\mathcal{R}=8,10,\overline{10},\cdots$ and
$Y,T,T_{3}$ are the corresponding hypercharge, isospin, and its third
component, respectively. Right hypercharge $Y^{\prime}$ is constrained
to be unity for the physical spin states for which $J$ and $J_{3}$ are
spin and its third component. Note that under the action of left
(flavor) generators
$\hat{T}_{\alpha}=-D_{\alpha\beta}^{(8)}\hat{J}_{\beta}$
$\psi_{B_{\mathcal{R}}}$ transforms 
like a tensor in representation $\mathcal{R}$, while under the right
generators $\hat{J}_{\alpha}$ like a tensor in $\mathcal{R}^{\ast}$ rather
than $\mathcal{R}$. This is the reason why operators like the one multiplied
by $v_{2}$ in Eq.(\ref{Eq:mu}) have different matrix elements for the decuplet
(which is spin $3/2$) and antidecuplet (which is spin $1/2$). The other two
operators multiplied by $v_{1,3}$ have the same matrix elements between
decuplet and antidecuplet states.

Using Eq.(\ref{Eq:wave_f}) and the known formulae~\cite{KW} for the
action of 
$V_{\pm}=\hat{J}_{4}\pm i\hat{J}_{5},\;U_{\pm}=\hat{J}_{6}\pm i\hat{J}_{7}$
\[%
\begin{array}
[c]{ccccc}%
U_{+} & \nwarrow &  & \nearrow & V_{+}\\
&  &  &  & \\
V_{-} & \swarrow &  & \searrow & U_{-}%
\end{array}
\]
on the antidecuplet ($\mathcal{R}=(0,q+2)$ with $q=1$) states,
\begin{figure}
[ptb]
\begin{center}
\includegraphics[
trim=0.000000in 0.000000in -0.001785in 0.000000in,
height=2.2087in,
width=2.3756in
]%
{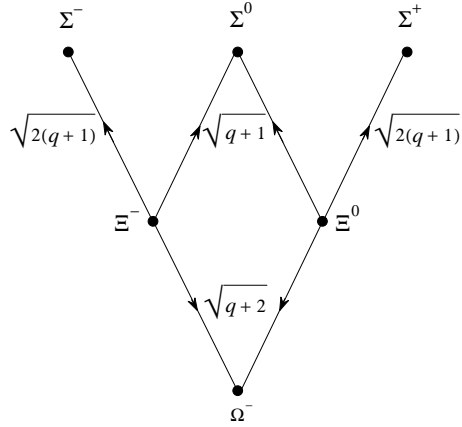}%
\caption{Action f $U_{\pm}$ and $V_{\pm}$ operators on states determining the
spin of antidecuplet ($q=1$) in $\mathcal{R}^{\ast}=(q+2,0)$. }%
\label{fig1}%
\end{center}
\end{figure}
%EndExpansion
we can easily calculate the collective matrix elements of the magnetic moments
of the baryon antidecuplet in Eq.(\ref{Eq:matrix_e}). The matrix elements of
Eq.(\ref{Eq:matrix_e}) are expressed just in terms of SU(3) Clebsch-Gordan
coefficients \cite{KW}. For example, the first term of the magnetic moment in
Eq.(\ref{Eq:mu}) in the case of $\Theta^{+}$ can be calculated as follows:
\begin{align}
&  \int dR\,\psi_{\Theta^{+}}^{\ast}(R)D_{Q3}^{(8)}(R)\psi_{\Theta^{+}%
}(R)=10\int dR\,D_{2,0,0;1,1/2,-1/2}^{(\overline{10})\ast}(R)D_{Q3}%
^{(8)}(R)D_{2,0,0;1,1/2,-1/2}^{(\overline{10})}(R)\nonumber\\
&  =\frac{1}{2}\left[  \left(  \left.
\begin{array}
[c]{cc}%
8 & \overline{10}\\
010 & 200
\end{array}
\right|
\begin{array}
[c]{c}%
\overline{10}\\
200
\end{array}
\right)  +\frac{1}{\sqrt{3}}\left(  \left.
\begin{array}
[c]{cc}%
8 & \overline{10}\\
000 & 200
\end{array}
\right|
\begin{array}
[c]{c}%
\overline{10}\\
200
\end{array}
\right)  \right]  \left(  \left.
\begin{array}
[c]{cc}%
8 & \overline{10}\\
010 & 1\frac{1}{2}-\frac{1}{2}%
\end{array}
\right|
\begin{array}
[c]{c}%
\overline{10}\\
1\frac{1}{2}-\frac{1}{2}%
\end{array}
\right)  \nonumber\\
&  =-\frac{1}{24}.
\end{align}
We can compute all relevant matrix elements in a similar manner. Having
scrutinized the results, we find the following simple expression:
\begin{equation}
\mu_{B_{\overline{10}}}=-\frac{1}{12}\,\left(  v_{1}+\frac{5}{2}v_{2}%
-\frac{1}{2}v_{3}\right)  Q_{B_{\overline{10}}}\,J_{3},\label{m0B10}%
\end{equation}
where $Q_{B_{\overline{10}}}$ is the charge of the antidecuplet expressed by the
Gell-Mann--Nishijima relation:
\begin{equation}
Q_{B_{\overline{10}}}=T_{3}+\frac{Y}{2}.
\end{equation}
$J_{3}$ is the corresponding third component of the spin.

In order to fit the parameters $v_{i}$, it is convenient to introduce two
parameters consisting of $v_{1}$, $v_{2}$ and $v_{3}$:
\begin{equation}
v=\frac{1}{60}\left(  v_{1}-\frac{1}{2}v_{2}\right)  ,~~w=\frac{1}{120}%
\;v_{3}. \label{vw}%
\end{equation}
In Ref.~\cite{Kim:1997ip} the octet and decuplet magnetic moments were
expressed as follows:
\begin{align}
\mu_{p}  &  =\mu_{\Sigma^{+}}=-8v+4w,\nonumber\label{Eq:all_mom}\\
\mu_{n}  &  =\mu_{\Xi^{0}}=6v+2w,\nonumber\\
\mu_{\Lambda}  &  =-\mu_{\Sigma^{0}}=3v+w,\nonumber\\
\mu_{\Sigma^{-}}  &  =\mu_{\Xi^{-}}=2v-6w,\nonumber\\
\mu_{B_{10}}  &  =\frac{15}{2}\left(  -v+w\right)  \,Q_{B_{10}}.
\end{align}
which are in fact the well-known SU(3) formulae for the magnetic moments.

The magnetic moments of the baryon antidecuplet (\ref{m0B10}) can be
rewritten as: 
\begin{equation}
\mu_{B_{\overline{10}}}=\left[  \frac{5}{2}\left(  -v+w\right)  -\frac{1}%
{8}v_{2}\,\right]  Q_{B_{\overline{10}}}.\label{Eq:final1}%
\end{equation}
Interestingly, the magnetic moments of the antidecuplet are different from
those of the decuplet by the second term in Eq.(\ref{Eq:final1}). The
factor three difference in the first term between
Eq.(\ref{Eq:all_mom}) and Eq.(\ref{Eq:final1}) is due to the fact that
the baryon antidecuplet has spin $1/2$, while the decuplet has $3/2$.
Additional term including $v_2$ appears due to the different action of
the second term in Eq.(\ref{Eq:mu}) on spin 1/2 and 3/2 states. 

Using Eq.(\ref{Eq:final1}), we can derive the sum rules which are similar to
the generalized Coleman and Glashow sum rules~\cite{Coleman:1961jn} in the
chiral limit:
\begin{align}
\mu_{\Sigma_{\overline{10}}^{0}}  &  =\frac{1}{2}\left(  \mu_{\Sigma
_{\overline{10}}^{+}}+\mu_{\Sigma_{\overline{10}}^{-}}\right)  ,\nonumber\\
\mu_{\Xi_{3/2}^{+}}+\mu_{\Xi_{3/2}^{--}}  &  =\mu_{\Xi_{3/2}^{0}}+\mu
_{\Xi_{3/2}^{-}},\nonumber\\
\sum\mu_{B_{\overline{10}}}  &  =0. \label{Eq:ColGl}%
\end{align}

As discussed in Ref.~\cite{Kim:1997ip}, there are different ways to fix the
parameters $v$ and $w$ by using the experimental data of the octet magnetic
moments. Here, we simply fit the proton and neutron magnetic moments (fit I):
\begin{equation}%
\begin{array}
[c]{llllr}%
v & = & (2\mu_{\mathrm{n}}-\mu_{\mathrm{p}})/20 & = & -0.331,\\
w & = & (4\mu_{\mathrm{n}}+3\mu_{\mathrm{p}})/20 & = & 0.037,
\end{array}
\label{Eq:fitI}%
\end{equation}
and use the following ''average'' values (fit II):
\begin{equation}%
\begin{array}
[c]{ccrcc}%
v & = & \left(  2\mu_{\mathrm{n}}-\mu_{\mathrm{p}}+3\mu_{\Xi^{0}}+\mu_{\Xi
^{-}}-2\mu_{\Sigma^{-}}-3\mu_{\Sigma^{+}}\right)  /60 & = & -0.268,\\
w & = & \left(  3\mu_{\mathrm{p}}+4\mu_{\mathrm{n}}+\mu_{\Xi^{0}}-3\mu
_{\Xi^{-}}-4\mu_{\Sigma^{-}}-\mu_{\Sigma^{+}}\right)  /60 & = & 0.060.
\end{array}
\label{Eq:mean}%
\end{equation}
to fix parameters $v$ and $w$. It was shown in Ref.~\cite{Kim:1997ip} that
combinations of Eq.(\ref{Eq:mean}) are independent of the linear corrections
due to the nonzero strange quark mass $m_{\mathrm{s}}$.  Thus, fit II
is also valid when the SU(3)-symmetry breaking is taken
into account, while fit I will be changed
by the corrections of order $\mathcal{O}(m_{\mathrm{s}})$. The results
of these fits are listed in Table I. 
\[%
\begin{array}
[c]{crrrr}
& \text{exp.} & \text{fit I} & \text{fit II} & \text{$\chi$QSM}\\\hline
p & 2.79 & \text{input} & 2.39 & 2.27\\
n & -1.91 & \text{input} & -1.49 & -1.55\\
\Lambda & -0.61 & -0.96 & -0.74 & -0.78\\
\Sigma^{+} & 2.46 & 2.79 & 2.38 & 2.27\\
\Sigma^{0} & (0.65) & 0.96 & 0.74 & 0.78\\
\Sigma^{-} & -1.16 & -0.89 & -0.90 & -0.71\\
\Xi^{0} & -1.25 & -1.91 & -1.49 & -1.55\\
\Xi^{-} & -0.65 & -0.89 & -0.90 & -0.71\\\hline
\Delta^{++} & \;4.52 & 5.52 & 4.92 & 4.47\\
\Omega^{-} & -2.02 & -2.76 & -2.46 & -2.23\\\hline
\Theta^{+} & ? & 0.30 & 0.20 & 0.12
\end{array}
\]
We see that the quality of these fits is rather poor reaching in its
worst case about $25\%$ accuracy, which indicates the importance of the 
SU(3)-symmetry breaking corrections. 

For the baryon antidecuplet we cannot predict the magnetic moments
unambigously, because they depend on the new parameter $v_{2}$,
unknown from nonexotic baryons.  Therefore, we have to resort to model 
calculations (\ref{spec}) of parameters $v_{1,2,3}$ which can be found
in the literature~\cite{Goeke:fk,Wakamatsu:1990ud,Blotz:1992pw}:%
\begin{equation}
v=-0.264,\;w=0.029,\;v_{2}\sim 5.\label{Eq:NJL}%
\end{equation}
We see that these parameters are in a rough agreement with the
model-independent analysis. It is therefore reasonable to assume that
$v_{2}$ is approximately equal to $5$. With this assumption we get the
predictions for the $\Theta^{+}$ magnetic moment which are listed in
Table I . We see that in all three cases our prediction for $\Theta^{+}$
magnetic moment is very small, almost order of magnitude smaller than
the magnetic moments for charge $Q=1$ particles both in the octet and
decuplet. Even if we take into account a $25\%$ error, a typical error of the
SU(3) symmetry fits, we still face yet another facet of exoticness
of the baryon antidecuplet: excessively small magnetic moments. 

The smallness of the $\Theta^{+}$ magnetic moment is due to the cancellation
of two terms in Eq.(\ref{Eq:final1}) $(-v+w)$ (which is positive) and
$v_{2}$. This is very similar to the cancellation which occurs in the case of
the $\Theta^{+}$ width \cite{Diakonov:1997mm}.  For these three fits discussed
above we have:%
\begin{equation}%
\begin{array}
[c]{ccc}%
\mu_{\Theta^{+}}=0.92-v_{2}/8 &  & \text{fit I,}\\
\mu_{\Theta^{+}}=0.82-v_{2}/8 &  & \text{fit II,}\\
\mu_{\Theta^{+}}=0.75-v_{2}/8 &  & \text{$\chi$QSM.}%
\end{array}
\end{equation}
Therefore, even if the model prediction for $v_{2}$ has $50\%$ error, the
largest value for $\mu_{\Theta^{+}}$ which we can get is the one from
fit I for $v_{2}=2.5$, $i.e.$ $\mu_{\Theta^{+}}=0.61$, still a pretty
small number. 

The expression of Eq.(\ref{Eq:mean}) leads to the additional sum rule for the
magnetic moments of the baryon antidecuplet:
\begin{equation}
\mu_{\Theta^{+}}-\mu_{\Xi_{3/2}^{--}}=\frac{1}{4}(2\mu_{\mathrm{p}}%
+\mu_{\mathrm{n}}+\mu_{\Sigma^{+}}-\mu_{\Sigma^{-}}-\mu_{\Xi^{0}}-2\mu
_{\Xi^{-}}), \label{Eq:theta}%
\end{equation}
which is very similar to Eq.(1) in Ref.~\cite{Kim:1997ip}.

\textbf{3.} If in the $\chi$QSM one artificially sets the soliton size
$r_{0}\rightarrow0$, then the model reduces to the free valence quarks which,
however, ''remember'' the soliton structure. In this limit, many
quantities, for example the axial-vector couplings, are given as
ratios of the group-theoretical factors \cite{limit}. In the case of
magnetic moments the pertinent expressions are given as a product of
the group-theoretical factor and the model-dependent integral which
we shall in what follows denote by $K$~\cite{paradox}. 

Constants $v_{1,2,3}$ entering Eq.(\ref{Eq:mu}) are expressed in terms of the
inertia parameters in the following way%
\begin{equation}
v_{1}=M_{0}-\frac{M_{1}^{(-)}}{I_{1}^{(+)}},\quad v_{2}=-2\frac{M_{2}^{(+)}%
}{I_{2}^{(+)}},\quad v_{3}=-2\frac{M_{1}^{(+)}}{I_{1}^{(+)}}.
\end{equation}
For the soliton size $r_{0}\rightarrow0$ we have \cite{paradox}:%
\begin{equation}
M_{0}\rightarrow-2K\ ,\quad\frac{M_{1}^{(-)}}{I_{1}^{(+)}}\rightarrow
\frac{4}{3}K\ ,~~~~\frac{M_{2}^{(+)}}{I_{2}^{(+)}}\rightarrow-\frac{4}%
{3}K~~~~~~~\frac{M_{1}^{(+)}}{I_{1}^{(+)}}\rightarrow-\frac{2}{3}K\ ,
\end{equation}
which give%
\begin{equation}
v=-\frac{7}{90}K,~~w=\frac{1}{90}K,\quad v_{3}=\frac{4}{3}K,
\end{equation}
yielding the magnetic moments of the proton and neutron as follows:
\begin{equation}
\mu_{p}=\frac{2}{3}K,\quad\mu_{n}=-\frac{4}{9}K. \label{Eq:Knp}%
\end{equation}
Hence, the ratio of the proton magnetic moment to the neutron one
takes the value from the nonrelativistic quark model:
\begin{equation}
\frac{\mu_{p}}{\mu_{n}}=-\frac{2}{3}.
\end{equation}
For antidecuplet magnetic moments we get%
\begin{equation}
\mu_{B_{\overline{10}}}=-\frac{1}{3}KQ_{B_{\overline{10}}}%
\end{equation}
which differs in sign from the phenomenological value of Table I (note that
$K$ is positive in view of Eq.(\ref{Eq:Knp})).  This is caused by the large
value of $v_{3}$ in the quark-model limit.  It would be interesting to see what
the quark models discussed in the literature give for the magnetic moment of
$\Theta^{+}$.

\textbf{4.} In the present work, we determined the magnetic moments of the
baryon antidecuplet in a ``model independent'' analysis, based on the chiral
quark-soliton model in the chiral limit.  Starting from the collective
operators with dynamical parameters fixed by experimental data, we
were able to obtain the magnetic moments of the baryon antidecuplet up
to one unknown constant which we have estimated from the model
calculations of
Refs.\cite{Goeke:fk,Wakamatsu:1990ud,Blotz:1992pw}. The expression for
the magnetic moments of the antidecuplet is different from those of
the baryon decuplet.  We found that the magnetic moment of
$\mu_{\Theta^{+}}$ is about $0.2\sim 0.3\,\mu_{N}$ which is surprisigly small
and is in line with the recent result of Ref.~\cite{summag}.

In the present letter, we have worked in the chiral limit.  The SU(3)-symmetry
breaking effects will definitely make the magnetic moments of the baryon
antidecuplet deviate from those of the present paper.  There are two different
sources of the SU(3)-symmetry breaking effects: one comes from the collective
operator, the other arises from the fact that the collective wave
functions of the baryon antidecuplet are mixed with the octet and
eikosiheptaplet representations.  Moreover, nonanalytical symmetry
breaking effects are of  importance \cite{chiloop}.  The effect
of the SU(3)-symmetry breaking on the magnetic moments of the
antidecuplet baryons is under investigation, however, our previous
experience shows that these effects do not exceed $25\%$.  Therefore,
we expect that the overall conclusion that the antidecuplet magnetic
moments are small will remain unchanged. 

\section*{Acknowledgments}

H.-Ch.K is grateful to J.K. Ahn (LEPS collaboration), K. Goeke,
A. Hosaka, M.V. Polyakov and I.K. Yoo (NA49 collaboration) for
valuable discussions.  The present work is supported by the KOSEF
grant R01\--2001\--00014 (H.-Ch.K.) and by the Polish State Committee
for Scientific Research under grant 2 P03B 043 24 (M.P.). This
manuscript has been authored  under Contract No. DE-AC02-98CH10886
with the U. S. Department of Eneregy.

\end{document}